\documentclass[fleqn,usenatbib]{mnras}

\usepackage{newtxtext,newtxmath}

\usepackage[T1]{fontenc}

\DeclareRobustCommand{\VAN}[3]{#2}
\let\VANthebibliography\thebibliography
\def\thebibliography{\DeclareRobustCommand{\VAN}[3]{##3}\VANthebibliography}

\usepackage{graphicx}	
\usepackage{amsmath}	
\usepackage{xspace}
\usepackage{pdflscape}
\usepackage{threeparttable}

\makeatletter
\newcommand\thefontsize[1]{{#1 The current font size is: \f@size pt\par}}
\makeatother

\title[Int-z ULIRGs between MS and SB]{Intermediate-redshift ultra-luminous infrared galaxies as transitioning objects between main sequence and starburst galaxies}

\author[J.-H. Kim et al.]{
Jee-Ho Kim,$^{1}$\thanks{E-mail: jee-ho.kim@physics.ox.ac.uk}
Dimitra Rigopoulou,$^{1,2}$
Miguel Pereira-Santaella,$^{3}$
Laia Barrufet,$^{4}$
\newauthor
Almudena Alonso-Herrero,$^{5}$
Ismael Garc\'ia-Bernete,$^{5}$
Chris Pearson$^{6}$
\\
$^{1}$Department of Physics, University of Oxford, Keble Road, Oxford OX1 3RH, UK\\
$^{2}$School of Sciences, European University Cyprus, Diogenes street, Engomi, 1516 Nicosia, Cyprus\\
$^{3}$Instituto de F\'isica Fundamental, CSIC-IFF, Calle Serrano 123, 28006 Madrid, Spain\\
$^{4}$Institute for Astronomy, University of Edinburgh, Royal Observatory, Edinburgh EH9 3HJ, UK\\
$^{5}$Centro de Astrobiolog\'ia (CAB), CSIC-INTA, Camino Bajo del Castillo s/n, E-28692, Villanueva de la Ca\~nada, Madrid, Spain\\
$^{6}$STFC Rutherford Appleton Laboratory, Harwell Campus, Didcot, OX11 0QX, UK
}

\date{Accepted XXX. Received YYY; in original form ZZZ}

\pubyear{\the\year{}}

\begin{document}
\label{firstpage}
\pagerange{\pageref{firstpage}--\pageref{lastpage}}
\maketitle

\begin{abstract}
With their star formation efficiencies (SFE) and star formation rates (SFR) falling between normal star-forming galaxies and starbursts (SBs), ultra-luminous infrared galaxies (ULIRGs) at intermediate redshifts ($0.2<z<1$) have been postulated to be good probes of galaxies transitioning between these two regimes.
Here, we present new Atacama Large Millimetre Array (ALMA) observations of a representative sample of 22 ULIRGs at redshifts $0.46<z<0.81$ evenly distributed across dust temperature ranges of $25<T_{\rm dust}<50$~K and spanning infrared luminosities of $10^{12.0}<L_\text{IR}<10^{12.7}~L_\odot$.
We detected CO(2--1) emission in 14 of these galaxies (64\%).
The SFE of these 14 targets are all similar, corresponding to depletion timescales on the order of hundreds of Myr.
By employing two different versions of the main sequence relation of galaxies --- one based on the specific SFR (sSFR) and one based on directly observable parameters (CO and IR luminosities) --
we statistically confirm that
the present sample of intermediate redshift ULIRGs
indeed lies in the transitional region between MS and SB galaxies.
In addition, we establish that the SFE of our sample of ULIRGs is a strong indicator of 
the position of a galaxy relative to the MS and the SB branches.
Based on the observed morphologies and star-forming properties, we suggest that minor mergers provide a viable  mechanism for temporarily enhancing the SFR and SFE of intermediate-redshift ULIRGs, although other evolutionary phases and pathways cannot be excluded.

\end{abstract}

\begin{keywords}
galaxies: star formation -- galaxies: starburst -- galaxies: evolution
\end{keywords}



\section{Introduction}

Ultra-luminous infrared galaxies (ULIRGs) are defined by their high luminosity at infrared wavelengths: $10^{12} < L_\text{IR} < 10^{13}~L_{\odot}$, where $L_\text{IR}$ is taken over 8--1000~$\mu$m \citep{1988Sanders}. Their lower luminosity counterparts, the luminous infrared galaxies (LIRGs), span the region just below at $10^{11} < L_\text{IR} < 10^{12}~L_{\odot}$.  
Both ULIRGs and LIRGs (henceforth referred to as U/LIRGs) are characterised by large reservoirs of molecular gas used to form stars \citep[e.g.,][]{2004GaoSolomon}.
Indeed, U/LIRGs have been found to have star formation rates (SFRs) on the order of 10--1000 $\text{M}_\odot ~ \text{yr}^{-1}$ e.g. \citep{2000Rowan-Robinson,2004Zheng}. Additionally, they contribute roughly half the total SFR density (SFR per comoving volume) above $z=1$ \citep{2005LeFloc'h,2011Magnelli,2012Casey}. As ULIRGs are responsible for most of the Universe's star formation rate at cosmic noon ($\sim$2--3 billion years after the Big Bang), studying the properties of this population of highly star-forming galaxies across redshift is imperative to understanding galaxy evolution.

It is widely accepted that a tight correlation ($<0.34$ dex scatter) exists between the SFR and stellar masses ($M_\star$) of galaxies, which defines the main sequence (MS) of star formation \citep{2004Brinchmann,2012Whitaker,2015Schreiber,2023Popesso}. The specific star formation rate $\text{sSFR} = \text{SFR}/M_\star$ positions a galaxy in the MS parameter space. Star-forming galaxies appear to have a bimodal distribution by sSFR, roughly splitting the population into main sequence galaxies and starburst galaxies (SB) \citep{2011Rodighiero,2012Sargent}. Many galaxies stay tightly aligned with the main sequence relation, hence their 
classification as MS galaxies. Starburst galaxies, however, are defined 
as having 
an sSFR at least four times greater than the MS relation, and are located higher in the parameter space \citep{2011Rodighiero,2018Garrett,2015Schreiber}.

Another 
well-established
correlation in galaxy studies is the Kennicutt-Schmidt relation, connecting the gas surface density of a galaxy and its SFR surface density as $\Sigma_\text{SFR} \propto \Sigma_\text{gas}^N$ \citep{1959Schmidt,1998Kennicutt}. The Kennicutt-Schmidt relation has a slightly lowered exponential factor $N$ when considering only the molecular gas surface density $\Sigma_\text{mol}$ compared to total gas surface density \citep{2008Bigiel,2011Schruba,2013Leroy,2019delosReyes}. The molecular gas content and SFR value of a galaxy are crucial in constraining its star formation efficiency $\text{SFE} = \Sigma_\text{SFR}/\Sigma_\text{mol}$ and depletion time, defined as $\tau = 1/\text{SFE} = \Sigma_\text{mol}/\Sigma_\text{SFR}$. The SFE and depletion time indicate how quickly a galaxy converts its molecular gas into stars 
\citep{1999Young,2013Leroy}. The notable increase in the SFE of SBs compared to MS galaxies imply their star formation processes 
are probably 
driven by different processes \citep{2020Tacconi}. Generally, it is accepted that MS galaxies undergo steady gas accretion, whereas SB galaxies tend to experience more dramatic events such as major mergers, resulting in episodes of intense SF activity \citep{1987ArmusHeckmanMiley,2007Noeske,2010Daddi,2019Pearson}. 

Based on observations of a small sample of 
intermediate-$z$ ($0.2<z<1$) ULIRGs, \citet{2022Hogan} found 
that  they tend to fall in between the locus defined by MS and SB galaxies, making them an ideal probe for the MS-SB transition region. 
These galaxies  
are distinct from SBs as they have lower SFEs, cooler dust temperatures and do not appear to be major mergers 
\citep{2014Rigopoulou,2014Magdis,2019Pereira-Santaella,2017Lee}. However, some SB-like characteristics (e.g., high SFEs) have been found among warmer ULIRGs at this redshift \citep{2011Combes,2013Combes}.  
Therefore, understanding the properties of intermediate-$z$ ULIRGs could potentially provide the best insight into understanding the MS-SB transition.

These intermediate-$z$ ULIRGs are distinct from their lower redshift
counterparts which are firmly located in the SB regime above the main sequence of galaxies.
Such low-$z$ ($z<0.2$) ULIRGs are more rare, 
having a low space density \citep[e.g.,][]{1988Sanders}.
Including the more populous LIRGs, local U/LIRGs as a whole 
share many SB characteristics.
In addition, they tend to have more compact emission, with a high fraction containing deeply obscured nuclei \citep{2011Iwasawa,2017Ricci,2018Torres-Alba,2021Ricci,2022Bellocchi}.
Their high SFR have been linked to interactions and mergers \citep{1987Armus,1990MelnickMirabel,2001Farrah}. These galaxies also have a high SFE that can be roughly an order of a magnitude higher than local disks \citep{2004GaoSolomon}. 

In contrast, high-$z$ ULIRGs ($z > 1$) lie closer to the main sequence, with a higher fraction ($\approx$ 50\%) of objects showing no sign of interaction \citep{2012Kartaltepe}.
Their physical sizes are generally larger than their low-$z$ counterparts, following a trend of increasing size with redshift \citep{2010Genzel,2010DaddiB,2019Pereira-Santaella,2022Hogan}. The dust temperatures are also cooler in these higher-$z$ ULIRGs \citep{2010Muzzin,2018Clements}.
Comparisons between intermediate-$z$ ULIRGs and complements at other redshifts thus can offer broad insight into the mechanisms motivating the MS-SB distribution.

The aim of this paper is to derive a physical characterization of intermediate-$z$ ULIRGs and investigate the connection between galaxy morphology and SFE.
We organize the paper as follows: in Section~\ref{sec:sample_selection}, we present the sample selection and relevant properties; in Section~\ref{sec:data}, we present the new ALMA observations and the data reduction steps along with archival and previously published data used to supplement our work; in Section~\ref{sec:analysis}, we review the analytical methods applied to the data to derive physical characterizations of our sample; in Section~\ref{sec:discussion}, we discuss the results of our analysis and contextualize them within the broader literature; lastly, in Section~\ref{sec:conclusions}, we summarize our findings and conclude with final remarks.

A cosmology of $\Omega_m = 0.3$, $\Omega_\Lambda = 0.7$, and $H_0 = 70\text{~km~s}^{-1}\text{~Mpc}^{-1}$ is assumed.

\section{Sample selection \& properties} \label{sec:sample_selection}

\subsection{Sample selection}
The intermediate-$z$ sample presented in this study has been drawn from Herschel observations of the XMM-LSS and CDFS fields.
We used the multi-wavelength catalogues of the {\it Herschel} Extragalactic Legacy Project \citep[HELP;][]{2021Shirley}, which provide deblended fluxes derived with the XID+ technique for Herschel detections \citep{2017Hurley}. The XMM-LSS field covers 11~deg$^2$, and the CDFS field covers 0.33~deg$^2$. The HELP catalog includes data of 170 million objects in 23 extragalactic survey fields in the optical to mid-infrared, cross-matched across 51 public surveys. These two fields were chosen because they (i) are the largest southern {\it Herschel} fields with sufficient depth to detect all ULIRGs ($L_{\rm IR} > 10^{12}L_{\odot}$) with dust temperatures $25 < T_{\rm dust} < 50$ K up to $z=0.83$, (ii) contain a high number of secure spectroscopic redshifts, and (iii) benefit from broad ancillary coverage in $>22$ photometric bands from the optical to the submillimetre.

\begin{figure}
    \centering
    \includegraphics[]{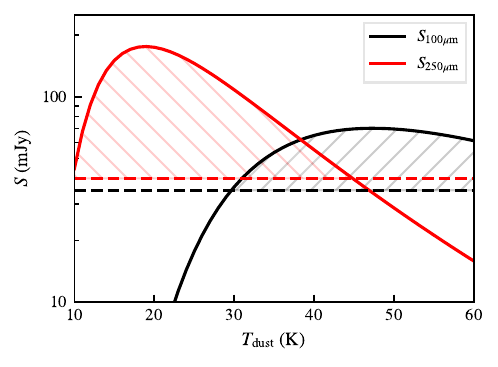}
    \caption{The two-band flux thresholds used for target selection at $L_\mathrm{IR}=10^{12}~L_\odot$ and $z=0.8$, shown as flux in mJy over $T_{\rm dust}$ in K. We use either $F_{100\mu m} > 35$~mJy or $F_{250\mu m} > 40$~mJy to obtain ULIRGs in a complete $T_{\rm dust}$ range. The inclusion of the $F_{100\mu m}$ band allows for consideration of warmer ULIRGs that may be too faint at $250~\mu m$. The cutoff for each threshold is represented with dashed lines, with the ULIRG selection regions filled with fainter hatches.}
    \label{fig:tdust_cut}
\end{figure}

We apply a two-band flux threshold to cull a comprehensive collection of galaxies from the XMM-LSS and CDFS fields.
Sources were initially selected to have either $F_{100\mu m} > 35$ mJy or $F_{250\mu m} > 40$~mJy, together with a reliable spectroscopic redshift. The combination of thresholds at two wavelengths ensures completeness across the full $T_{\rm dust}$ range, including warmer ULIRGs that may be faint at 250\,$\mu$m. Figure~\ref{fig:tdust_cut} depicts the two flux cuts as flux versus resulting dust temperature. This process yielded 3984 galaxies (3817 in XMM-LSS at $0.3 < z < 0.84$ and 167 in CDFS at $0.2 < z < 0.45$). We then performed SED fitting with \textsc{CIGALE} \citep{2019Boquien} for all selected sources.

\begin{figure}
	\centering
	\includegraphics[width=\columnwidth]{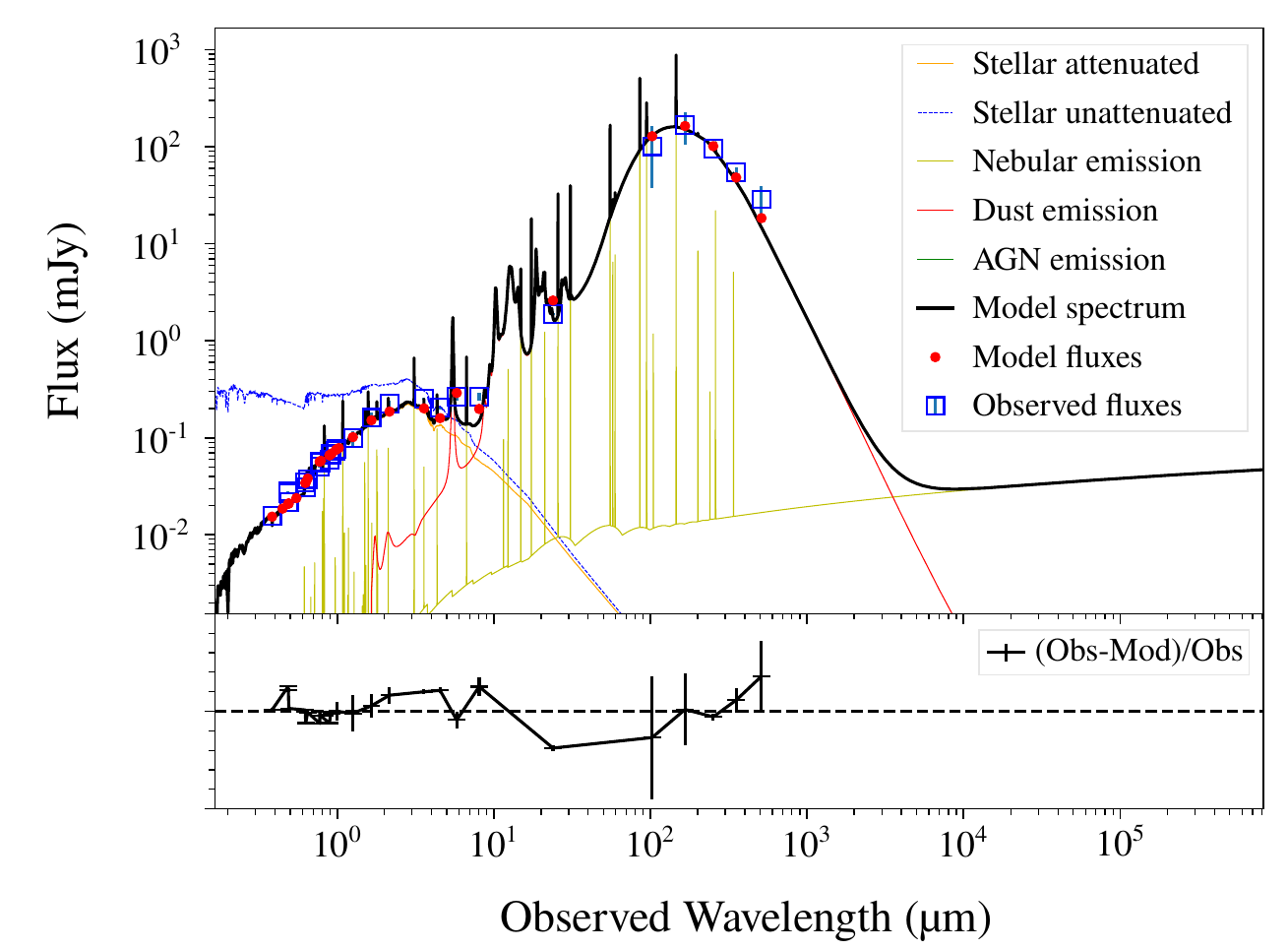}   
	\caption{The SED of one of the sources in the final sample selection (HELP\_J021654.489-035001.451) at $z=0.651$ and $L_{\rm IR}= 2.3 \times 10^{12}\,L_{\odot}$. 
	The squares are the 28 observed fluxes, whereas the filled circles are the fluxes from the models. 
	The thick solid line in the upper plot represents the overall model spectrum, including contributions from dust emission \citep{2007DraineLi}, AGN \citep{2006Fritz}, and stellar emission \citep{2003BruzualCharlot}. Model residuals are included in the lower subplot.}
	\label{SED_sample_selection}
\end{figure}

\begin{figure}
    \centering
    \includegraphics[]{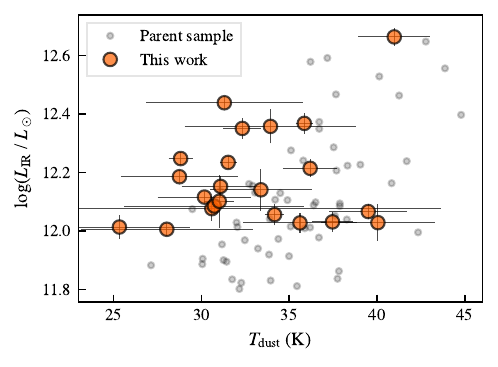}
    \caption{Intermediate-$z$ U/LIRGs, presented as $L_\text{IR}$ over $T_{\rm dust}$ in K. The smaller, fainter circular markers represent the parent sample for this study, prior to selecting the final targets for observations. The large, outlined circles indicate the final observed objects presented here.}
    \label{fig:LIR_tdust}
\end{figure}

Next, we narrow down the temperature-complete galaxy population to our final ULIRG sample by filtering by IR luminosity ($L_\text{IR}>8\times 10^{11} L_{\odot}$) and redshift ($0.3<z<0.83$).
Among the 3984 selected galaxies, 873 systems have $L_{\rm IR} > 8 \times 10^{11}\,L_\odot$, of which 524 exceed the ULIRG threshold. From this ULIRG parent sample, we defined a representative subset of 30 galaxies in the range $0.3 < z < 0.83$, uniformly sampling the redshift ($\sim$10 per bin: $0.3<z<0.48$, $0.48<z<0.66$, $0.66<z<0.83$), $L_{\rm IR}$, and $T_{\rm dust}$ (see Figure~\ref{SED_sample_selection} for a representative ULIRG SED).
We did not target galaxies in the lowest bin as galaxies in this redshift range were discussed by \citet{2022Hogan}, although their data are included in this study.
As such, our selection provides a representative sample of intermediate-$z$ ULIRGs that bridges the gap between warm, luminous starbursts at similar redshifts \citep{2011Combes} and colder, lower-luminosity main sequence galaxies \citep{2017Lee}. Figure~\ref{fig:LIR_tdust} presents our final observed sample of 22 $0.46 < z < 0.81$ ULIRGs compared to the larger parent sample as $L_\text{IR}$ over $T_\text{dust}$.

Although this selection was designed to provide representative coverage in $L_\text{IR}$, $T_\text{dust}$, and redshift within the ULIRG parent population, it should not be interpreted as an unbiased selection in all star-formation properties. In particular, the $L_\text{IR}$ requirement preferentially selects galaxies with high SFRs. Furthermore, analyses involving $L_\text{CO}'$ are necessarily affected by the CO sensitivity of the ALMA observations. We discuss the consequences of these selection effects for the inferred SFE distribution in Section~\ref{sec:disc-sfe}.

\subsection{The presence of AGN in our sample ULIRGs}

Based on our SED fitting, we infer the incidence 
of AGN in our sample.
Of the 30 initially selected objects, 
six of them required the presence of an AGN to fit their SED with CIGALE models. Only two of the 22 targets presented in this work have an AGN contribution in their SED fit. In addition, in the HELP master catalog, HELP\_J021654.489-035001.451 is 
flagged as an AGN based on optical spectra. 
Due to the low ($<$10--20\%) occurrence rate among these intermediate-$z$ ULIRGs, we do not consider the presence of the  AGN and the impact on our analysis further. We do point out, however, that our findings differ from the higher incidence of  AGN acitivity found among samples of local ULIRGs \citep{2009Veilleux,2010Nardini}. 

\section{Observational Data} \label{sec:data}

\subsection{ALMA observations and data reduction}

\begin{table*}
	\caption{The intermediate-$z$ ULIRG survey. From left to right: the galaxy names, the source RA and declination, the spectroscopic redshift of the galaxy, the on-source integration time, the synthesized beam size, and the sensitivity per channel. The redshift is pulled from the HELP catalog unless otherwise indicated.\label{tbl:obs_props}}
	\begin{threeparttable}[c]
		\centering
		\footnotesize
		\begin{tabular}{ccccccc}
		\hline
		Object                       & RA     & Dec    & $z_\mathrm{spec}$     & $t_\text{int}$ & Synthesized beam         & Sensitivity       \\
									&   (deg)     &  (deg)      &       & (hrs)          & (arcsec $\times$ arcsec) & (mJy beam$^{-1}$) \\ \hline
		HELP\_J021636.715-035351.753 & 34.153 & -3.898 & 0.572\tnote{\dag} & 0.9              & $0.237 \times 0.188$ & 0.341       \\
		HELP\_J021642.542-042154.994 & 34.177 & -4.365 & 0.583\tnote{\dag} & 0.7              & $0.240 \times 0.175$ & 0.342       \\
		HELP\_J021654.489-035001.451 & 34.227 & -3.834 & 0.651\tnote{\dag} & 0.8              & $0.236 \times 0.182$ & 0.339       \\
		HELP\_J021742.522-053700.041 & 34.427 & -5.617 & 0.678\tnote{\dag} & 0.8              & $0.243 \times 0.166$ & 0.287       \\
		HELP\_J021814.949-044915.295 & 34.562 & -4.821 & 0.756\tnote{\dag} & 0.8              & $0.280 \times 0.211$ & 0.367       \\
		HELP\_J021820.088-052155.477 & 34.584 & -5.365 & 0.720 & 0.9              & $0.264 \times 0.205$ & 0.378       \\
		HELP\_J021824.031-045304.606 & 34.600 & -4.885 & 0.458\tnote{\dag} & 0.7              & $0.228 \times 0.161$ & 0.415       \\
		HELP\_J021856.742-055437.350 & 34.736 & -5.910 & 0.633\tnote{\dag} & 0.8              & $0.254 \times 0.191$ & 0.371       \\
		HELP\_J021929.883-050900.724 & 34.875 & -5.150 & 0.671 & 0.8              & $0.242 \times 0.167$ & 0.301       \\
		HELP\_J021941.586-044426.974 & 34.923 & -4.741 & 0.760\tnote{\dag} & 0.8              & $0.225 \times 0.192$ & 0.363       \\
		HELP\_J022057.177-042611.207 & 35.238 & -4.436 & 0.480\tnote{*} & 0.7              & $0.225 \times 0.164$ & 0.346       \\
		HELP\_J022117.385-050752.719 & 35.322 & -5.131 & 0.802 & 1.5              & $0.228 \times 0.174$ & 0.362       \\
		HELP\_J022127.332-044349.647 & 35.364 & -4.730 & 0.615 & 0.8              & $0.206 \times 0.175$ & 0.336       \\
		HELP\_J022130.169-045002.352 & 35.376 & -4.834 & 0.809 & 1.5              & $0.227 \times 0.174$ & 0.363       \\
		HELP\_J022131.352-040809.328 & 35.381 & -4.136 & 0.621\tnote{\dag} & 0.8              & $0.252 \times 0.191$ & 0.410       \\
		HELP\_J022151.541-025349.283 & 35.465 & -2.897 & 0.690\tnote{\dag} & 0.9              & $0.255 \times 0.204$ & 0.317       \\
		HELP\_J022209.374-041137.227 & 35.539 & -4.194 & 0.616\tnote{\dag} & 0.8              & $0.206 \times 0.176$ & 0.338       \\
		HELP\_J022222.816-043143.814 & 35.595 & -4.529 & 0.797\tnote{*} & 1.5              & $0.228 \times 0.174$ & 0.362       \\
		HELP\_J022247.822-055432.792 & 35.699 & -5.909 & 0.687\tnote{\dag} & 0.8              & $0.249 \times 0.167$ & 0.345       \\
		HELP\_J022333.674-045627.366 & 35.890 & -4.941 & 0.826 & 1.5              & $0.232 \times 0.175$ & 0.449       \\
		HELP\_J022427.439-033006.637 & 36.114 & -3.502 & 0.798\tnote{\dag} & 0.8              & $0.249 \times 0.197$ & 0.341       \\
		HELP\_J022722.754-043350.854 & 36.845 & -4.564 & 0.640\tnote{\dag} & 0.8              & $0.232 \times 0.180$ & 0.407		\\ \hline
		\end{tabular}
		\begin{tablenotes}
			\item[\dag] Updated redshift based on ALMA spectra
			\item[*] Unreliable spectroscopic redshift; see Section~\ref{sec:analysis-comaps} for details
		\end{tablenotes}
	\end{threeparttable}
\end{table*}

In this section, we describe the ALMA observations of the 22 $0.46 < z < 0.81$ ULIRGs and the data reduction steps.

ALMA band 4 observations of the CO(J=2--1) emission line and continuum were taken using the interferometer's 12-m array (project 2023.1.00424.S, PI: D Rigopoulou). 
We select the J=2--1 transition of CO first for its brightness (compared to the J=1--0 transition), and also because it is more directly tracing the cold molecular gas linked to star formation. 
The observations took place over several nights between November 13, 2023 and November 30, 2023, with roughly 12--19 minutes of on-source integration time for each object. The synthesized beam for each target is included in Table~\ref{tbl:obs_props}, with the average resolution being $0.238\arcsec \times 0.181\arcsec$.

We reduced the data by running them through the standard pipeline in the Common Astronomy Software Applications (CASA) software \citep{2022CASA}. Calibration and flagging of the data were done according to the ALMA Quality Assurance process.

The line emission cubes were processed with the TCLEAN routine in CASA. A  Briggs weighting of 0.5 was selected. The pixel sizes of the cubes were chosen to be 
0.025\arcsec, and all cubes were cleaned by applying a 3$\sigma$ threshold. The native channel widths of the galaxies are roughly 8~km/s. 

All targets but one were either detected by ALMA as single sources or undetected.
Although HELP\_J021742.522-053700.041 appeared as a point source in Herschel images (with angular resolutions ranging from 5\arcsec--30\arcsec across the PACS and SPIRE instruments), our ALMA observations identified two sources within 2\arcsec of the centre coordinates in the CO data (Figure~\ref{fig:twosource}). We choose to focus our analysis on the southern source with stronger CO(J=2--1) emission.

\subsection{Suprime-Camera optical archival data}

Ancillary optical \textit{i}-band and \textit {g}-band images of the galaxies were taken from the Hyper Suprime-Cam Legacy Archive \citep[HSCLA;][]{2021Tanaka}. The HSC is an 870 Mega pixel camera developed for the 8.2-m Subaru Telescope in Maunakea, Hawai'i. HSC data are also included in the HELP catalog. The 8\arcsec cutouts of the galaxies were taken from coadded images in the database (Figure~\ref{fig:suprimecam}). \textit{i}-band images were selected by default, and when unavailable, \textit{g}-band data were pulled. All images, with the band specified, are compiled in the appendix (Figure~\ref{fig:suprimecam}). HELP\_J021856.742-055437.350 and HELP\_J022247.822-055432.792 were the only targets for which images were unavailable.

\subsection{Additional samples} \label{sec:add_samples}
To put our sample of intermediate-$z$ ULIRGs in context, we employ the following CO studies,
beginning with other populations at intermediate-$z$. The \citet{2022Hogan} sample preceding this work consists of five ULIRGs at $0.289<z<0.437$, three of which are interacting. \citet{2014Magdis} includes these targets in their larger survey of 17 $0.21<z<0.88$ U/LIRGs.
\citet{2017Lee} studied a sample 
of 20 LIRGs at similar redshift ($0.25<z<0.65$) but with cooler dust temperatures 
than those of the sample presented in this paper ($T_\text{dust}\lesssim 40$~K). 
Most of the \citet{2017Lee} sources appear to be isolated with disk-like morphologies. 
\citet{2011Combes} selected 30 warm ($T_\text{dust}\gtrsim 40$ K) ULIRGs at $0.2<z<0.6$. Two-thirds of the detected objects were found to be interacting. Our sample, in comparison, spans both the $L_\text{IR}$ and $T_\text{dust}$ ranges covered by the \citet{2017Lee} and \citet{2011Combes} studies.

We complemented these works with local samples.
The Valparaíso ALMA Line Emission Survey (VALES) combined with Atacama Pathfinder Experiment APEX/SEPIA observations forms a large sample of local LIRGs and MS galaxies at $z<0.2$ \citep{2018Cheng}. The population is a mix of isolated and interacting galaxies, intended to probe the MS-SB transition at low redshift.
The Physics of ULIRGs with the Multi-Unit Spectroscopic Explorer and ALMA (PUMA) project examines a representative sample of local ($z<0.165$) ULIRGs \citep{2021Pereira-Santaella}. Half of the 23 observed ULIRGs are interacting systems, and the other half are advanced mergers. CO data for PUMA targets were taken from \citet{2022Lamperti}.
\citet{2013Leroy} targeted 30 local disks from the HERA CO-line Extragalactic Survey (HERACLES).

The three high-$z$ samples referenced in this paper include mostly normal star-forming galaxies. 
The Pleateau de Bure High-z Blue Sequence Survey (PHIBSS) examines 52 MS galaxies at $1<z<3$, the majority of which have disk morphologies \citep{2013Tacconi}.
Additionally, we include 11 U/LIRGs ($0.8<z<2.6$) taken from the blind ALMA Spectroscopic Survey (ASPECS) project \citep{2016Decarli}. These targets were selected to be mostly non-interacting galaxies with half of them found to lie on the MS.
Finally, we include the sample of high-redshift galaxies presented in \citet{2009Iono}.

Table~\ref{tbl:surveys} summarizes these samples for convenient reference.

\section{Analysis} \label{sec:analysis}

The following sections present the methods used to process the ALMA CO line data of our ULIRG sample, convert them to physical quantitative properties, and finally,
to place them on the main sequence plot of galaxies and investigate their location 
relative to MS and SB populations from different galaxy samples at various redshifts using two different diagnostic plots.

\begin{figure*}
	\includegraphics{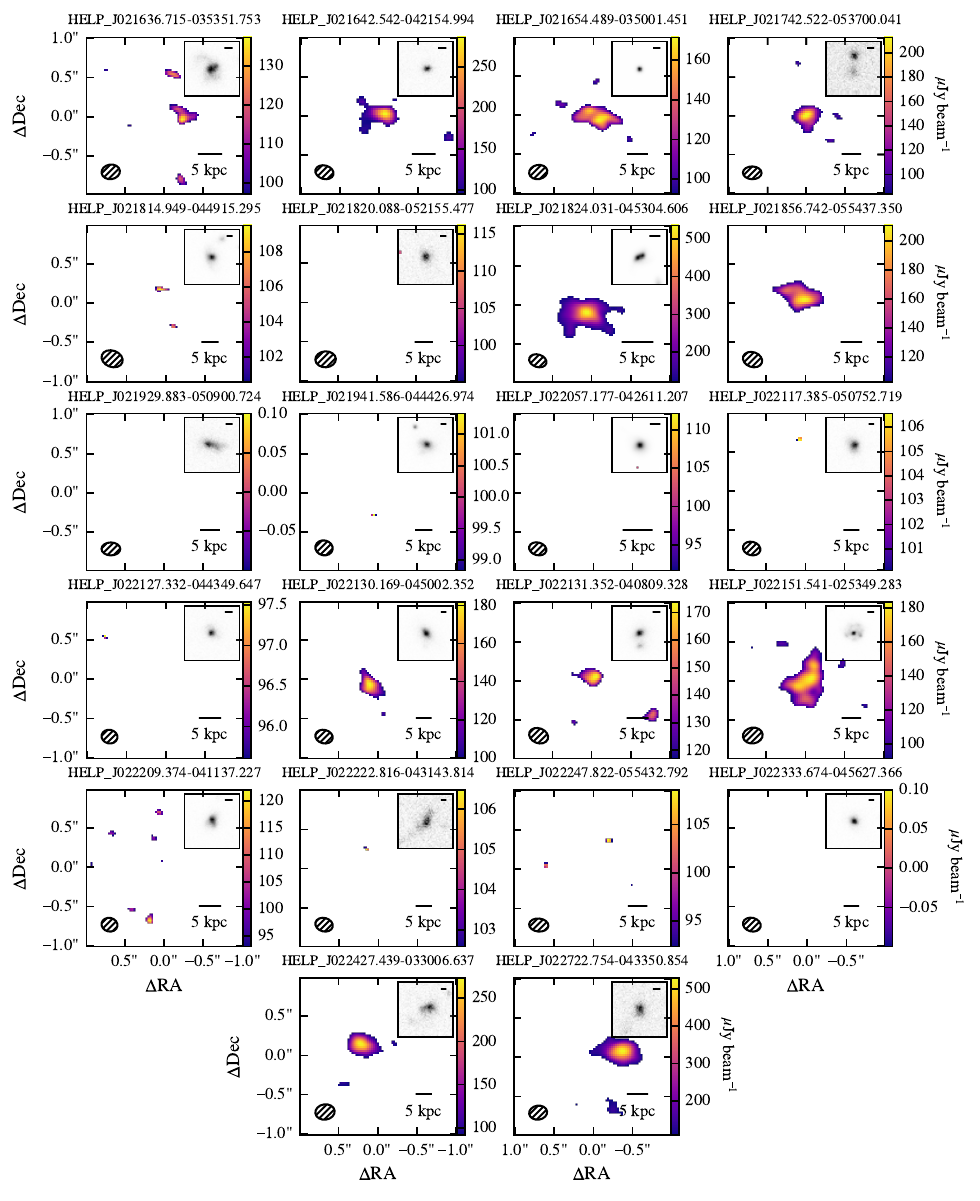}
    \caption{Moment 0 maps of the intermediate-$z$ sample. The bottom left of each image indicates the beam size, and the bar to the bottom right is a physical scale bar for 5~kpc. Overlaid in the top right corner are the Suprime-Cam images, which are enlarged in Figure~\ref{fig:suprimecam}. The bars in the upper right corner of the Suprime-Camera images measure 20~kpc.}
    \label{fig:3sig_mom0}
\end{figure*}

\subsection{CO(2--1) maps} \label{sec:analysis-comaps}

\begin{figure}
	\includegraphics{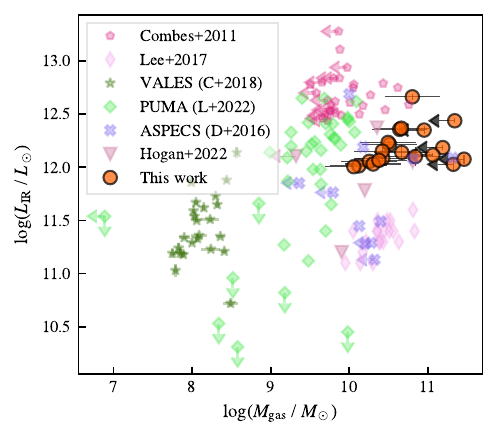}
    \caption{Total infrared luminosity vs. molecular gas mass. The intermediate-$z$ ULIRG sample are represented with large circles outlined in black. Other samples pulled from the literature include a smaller sample of intermediate-$z$ ULIRGs as triangles \citep{2022Hogan}, warm SB-like intermediate-$z$ ULIRGs as pentagons \citep{2011Combes}, cooler intermediate-$z$ LIRGs \citep{2017Lee} as diamonds, local star-forming galaxies as stars \citep{2018Cheng}, local ULIRGs as large diamonds \citep{2022Lamperti}, and high-$z$ U/LIRGs as crosses \citep{2016Decarli}.}
    \label{fig:luminosity_mass}
\end{figure}

\begin{table*}
	\caption{Physical properties of the intermediate-$z$ ULIRG sample, mostly derived from the SED fits. From left to right: the object name (shortened for brevity), dust temperature, effective radius, effective radius to average beam radius ratio, star formation rate, stellar mass, and infrared luminosity. The effective radius is only included for extended objects.\label{tbl:phys_props}}
	\begin{threeparttable}[c]
		\centering
		\footnotesize
		\begin{tabular}{lcccccc}
		\hline
		Object        & $T_\text{dust}$  & $R_\text{eff}$ &	$R_\text{eff}/R_\text{beam}$	& SFR                        & $M_\star$\tnote{\dag}                        & $\log L_\text{IR}$ \\
					& (K)              & (kpc)          &	& $\left(M_\odot~ \text{yr}^{-1}\right)$ & $\left( 10^{10}~M_\odot \right)$ &      $\left( L_\odot \right)$                                       \\ \hline
		HELP\_J021636 & $25.3 \pm 4.0$   & $2.64$  & $1.54$    & $95 \pm 16$     & $32.5 \pm 2.9$   & $12.01 \pm 0.04$ \\
		HELP\_J021642 & $34.17 \pm 0.55$ & $3.50$  & $2.04$    & $154 \pm 23$    & $7.77 \pm 0.72$  & $12.06 \pm 0.04$ \\
		HELP\_J021654 & $33.9 \pm 4.9$   & $4.15$  & $2.10$    & $257 \pm 34$    & $12.0 \pm 1.3$   & $12.36 \pm 0.06$ \\
		HELP\_J021742 & $35.86 \pm 0.50$ & $3.60$  & $1.77$    & $221 \pm 17$    & $8.37 \pm 0.59$  & $12.37 \pm 0.04$ \\
		HELP\_J021814 & $30.7 \pm 5.1$   & $<1.40$ & $<0.50$    & $89.1 \pm 5.5$  & $30.9 \pm 1.5$   & $12.09 \pm 0.02$ \\
		HELP\_J021820 & $28.7 \pm 3.3$   & -       & -         & $125 \pm 11$    & $21.7 \pm 1.1$   & $12.19 \pm 0.02$ \\
		HELP\_J021824 & $31.52 \pm 0.48$ & $3.63$  & $3.01$    & $248 \pm 17$    & $4.12 \pm 0.61$  & $12.23 \pm 0.02$ \\
		HELP\_J021856 & $33.4 \pm 2.9$   & $4.04$  & $1.99$    & $119 \pm 21$    & $23.2 \pm 3.7$   & $12.14 \pm 0.07$ \\
		HELP\_J021929 & $30.2 \pm 2.7$   & -       & -         & $135.7 \pm 6.8$ & $10.34 \pm 0.86$ & $12.12 \pm 0.02$ \\
		HELP\_J021941 & $39.5 \pm 2.2$   & -       & -         & $114 \pm 11$    & $9.1 \pm 1.8$    & $12.07 \pm 0.03$ \\
		HELP\_J022057 & $35.6 \pm 3.2$   & -       & -         & $118.9 \pm 8.3$ & $12.5 \pm 2.2$   & $12.03 \pm 0.03$ \\
		HELP\_J022117 & $40.0 \pm 3.2$   & -       & -         & $93 \pm 13$     & $10.9 \pm 1.3$   & $12.03 \pm 0.06$ \\
		HELP\_J022127 & $32.3 \pm 1.1$   & -       & -         & $218 \pm 20$    & $9.1 \pm 1.7$    & $12.35 \pm 0.04$ \\
		HELP\_J022130 & $31.3 \pm 4.5$   & $3.85$  & $1.55$    & $188.0 \pm 9.4$ & $66.5 \pm 3.3$   & $12.44 \pm 0.02$ \\
		HELP\_J022131 & $31.1 \pm 2.2$   & $2.86$  & $1.45$    & $111 \pm 12$    & $24.0 \pm 1.2$   & $12.15 \pm 0.04$ \\
		HELP\_J022151 & $36.2 \pm 1.5$   & $5.96$  & $2.56$    & $155 \pm 16$    & $14.4 \pm 2.7$   & $12.21 \pm 0.03$ \\
		HELP\_J022209 & $28.0 \pm 4.9$   & $<1.75$ & $<1.03$    & $73.3 \pm 4.7$  & $25.2 \pm 1.3$   & $12.01 \pm 0.02$ \\
		HELP\_J022222 & $28.81 \pm 0.69$ & -       & -         & $247 \pm 17$    & $12.24 \pm 0.73$ & $12.25 \pm 0.02$ \\
		HELP\_J022247 & $37.5 \pm 1.2$   & $<1.10$ & $<0.52$    & $115.5 \pm 7.4$ & $7.4 \pm 1.5$    & $12.03 \pm 0.03$ \\
		HELP\_J022333 & $31 \pm 13$      & -       & -         & $105 \pm 16$    & $20.7 \pm 1.1$   & $12.08 \pm 0.04$ \\
		HELP\_J022427 & $41.0 \pm 2.1$   & $4.69$  & $1.73$    & $482 \pm 66$    & $13.46 \pm 0.85$ & $12.66 \pm 0.03$ \\
		HELP\_J022722 & $31.02 \pm 0.85$ & $4.58$  & $2.40$    & $132 \pm 56$    & $10.7 \pm 1.3$   & $12.10 \pm 0.09$ \\ \hline
		\end{tabular}
		\begin{tablenotes}
			\item[\dag] The quoted $M_\star$ uncertainties correspond to the statistical uncertainties from the adopted SED-fitting procedure and do not include systematic uncertainties associated with the assumed stellar population model, star-formation history, attenuation prescription, or other modelling assumptions.
		\end{tablenotes}
	\end{threeparttable}
\end{table*}

We first generate the CO maps and line cubes
and extract the relevant CO line properties.

The CO(2--1) moment 0 maps were generated by integrating the line emission data along the spectral axis and performing a 3$\sigma$ cut. The maps are presented in Figure \ref{fig:3sig_mom0}. Only half of the 22 galaxies were found to be extended. For these extended objects, we estimate the effective radius $R_\text{eff}$ by summing the total area within a 2\arcsec$\times$2\arcsec radius aperture around the galaxy center after applying a 3$\sigma$ cut and calculate the radius, assuming a circular shape. The resulting $R_\text{eff}$ range from 1.10--5.96~kpc. Targets with an $R_\text{eff}$ less than half the beam's major axis radius were considered unresolved, and their $R_\text{eff}$ is marked as an upper limit in Table~\ref{tbl:phys_props}. Most of the sources appear to have CO gas concentrated in their cores, which are difficult to probe with the resolution of our observations. As such, we also include for reference the ratio of the effective radius to the average of the radii of the beam's major and minor axes in Table~\ref{tbl:phys_props}.

To extract the spectra, we first binned the line data by four channels to boost the signal, resulting in channel widths varying from 30--37~km/s. We applied a 0.5\arcsec radius mask around each galaxy, which we found sufficient enough to cover the extent of the emission. 
The spectra are shown in Figure~\ref{fig:spectra2} with the line emission shaded lightly.
Detected lines were fitted with 1-dimensional Gaussians (overplotted in the figure), from which the FWHM was calculated. These fits were limited by the binned channel widths.
We show the FWHMs derived from the spectra of each target in Table~\ref{tbl:line_props}.

Note that some non-detections could be the result of uncertain or misidentified redshifts. 
Most of the undetected galaxies have a spectroscopic redshift Quality Flag $Q\geq 3$, corresponding to a $>92\%$ reliability in the HELP catalogues \citep{2021Shirley}. One non-detection, HELP\_J022222.816-043143.814, had $Q=2$ with no reliability estimate available. Generally, spectroscopic redshifts with $Q \geq 3$ are 
considered reliable.
In the case of HELP\_J022057.177-042611.207, the spectroscopic redshift was taken from the SDSS catalog. However, the listed redshift of 0.480 in the HELP catalog does not match the spectroscopic redshift $z_\mathrm{spec}=0.315$ that was measured for the cross-matched SDSS target. Continuum was not detected for these or any of the other targets in this sample as well. Due to the ambiguous reliability of the measured redshifts of these last two targets, we remove them from consideration in further data analysis.

Once the data were processed, we then obtained the CO line luminosities of our targets. We converted the CO(2--1) flux density to line luminosity using the relation \citep{1997Solomon}:
\begin{equation}
	L_\text{CO}' = 3.25 \times 10^7 \cdot S_\text{CO} \Delta v \cdot \nu_\text{obs}^{-2} \cdot D_\text{L}^2 (1+z)^{-3}.
	\label{eqn:L'CO}
\end{equation}
$S_\text{CO} \Delta v$ is the integrated flux density in $\text{Jy~km~s}^{-1}$, $\nu_\text{obs}$ is the observed frequency in GHz, $D_\text{L}$ is the luminosity distance in Mpc, and $z$ is the redshift. We took $\nu_\text{obs}$ as the central frequency of the fitted Gaussian, and subsequently used that value to recalculate the redshifts of the galaxies, accurate to within a 0.23\% error of the original HELP values. 
Table~\ref{tbl:obs_props} indicates the updated redshifts.
The line emission width was taken as 3$\sigma$ on either side of $\nu_\text{obs}$. We calculated $S_\text{CO} \Delta v$ as the sum of the flux density over the line emission channels multiplied by the velocity channel width.  

For sources with no line detections and $Q\geq 3$, we provide upper limit estimates. 
We calculated a 3$\sigma$ upper limit for the integrated flux density following \citet{2022Hogan}. We first find the significant flux density per beam $3\sigma / \text{beam} = 3\sqrt{N}\delta v \sigma_\text{rms}$, where $N$ is the number of channels containing line emission, $\delta v$ is the channel width in $\text{km~s}^{-1}$ and $\sigma_\text{rms}$ is the noise per channel. The resultant $3\sigma / \text{beam}$ was then multiplied by the square root of the number of beams within the star-forming region $A_\text{CO}$. 
We assume the CO emission fully traces the star formation in each galaxy and 
estimate $A_\text{CO}$ as the number of beams within a circular region of 0.5\arcsec radius. We also set the upper limit of the line width as the broadest detected line in our sample, which is roughly 1500~km/s, and converted to number of channels for each non-detection.

The resulting values are summarized in Tables~\ref{tbl:phys_props} and \ref{tbl:line_props}.

\begin{figure}
	\includegraphics{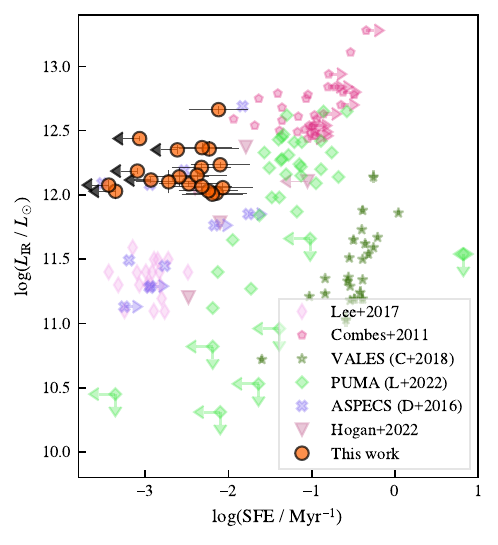}
    \caption{Infrared luminosity over star formation efficiency. The SFE of this sample (black-outlined circles) hover between cool MS-like LIRGs \citep{2017Lee} and warm SB-like ULIRGs \citep{2011Combes} at similar redshift. Local star-forming galaxies are included as stars for reference \citep{2018Cheng}, along with local ULIRGs as large diamonds \citep{2022Lamperti}, and high-$z$ U/LIRGs as crosses \citep{2016Decarli}.}
    \label{fig:luminosity_SFE}
\end{figure}

\subsection{Molecular gas mass} \label{sec:analysis-mgas}
From our derived $L_\text{CO}'$, we can compute physical  quantities for our sample for comparison with the literature. We begin with the most directly obtainable molecular gas mass.

The molecular gas mass of a galaxy can be derived with
\begin{equation}
	M_\text{gas} = \alpha L_\text{CO}',
	\label{eq:gas_mass}
\end{equation}
where $\alpha$ is the CO conversion factor. $\alpha_\text{MW} = 4.36 \pm 0.9~M_\odot(\text{K~km~s}^{-1}\text{~pc}^2)^{-1}$ is the Milky Way conversion factor, which can be applied to local MS galaxies \citep{1991SolomonBarrett,1996StrongMattox,2010Abdo,2013Bolatto,2015Genzel}. For local ULIRGs, $\alpha \sim 0.8~M_\odot(\text{K~km~s}^{-1}\text{~pc}^2)^{-1}$ is used \citep{1998DownesSolomon,2005Solomon}. More recent studies have found a dependence of $\alpha$ on the metallicity of a galaxy \citep{2011Leroy,2012Genzel,2014Remy-Ruyer,2019Coogan}. We therefore adopt the metallicity-based estimation of $\alpha_{\rm CO}$ for our sample 
from \citet{2012Genzel,2015Genzel}:
\begin{equation}
	\alpha_\text{CO(2--1)} = \alpha_\text{MW} \cdot \chi(Z) \cdot R_{12}.
	\label{eq:alphaCO}
\end{equation}
where $R_{12}=1.2$ is a correction factor accounting for the lower $J=2\text{--}1$ Rayleigh-Jeans brightness temperature compared to the ground transition \citep[and references therein]{2015Genzel}.

The metallicity dependence of equation~(\ref{eq:alphaCO}) comes from the fitting function
\begin{equation}
	\chi(Z) = 10^{ -1.27 \times \left( 12+\log(\text{O/H})-8.67 \right)},
	\label{eq:chi(Z)}
\end{equation}
where $12+\log(\text{O/H})$ is the galaxy's gas-phase oxygen abundance \citep{2004PettiniPagel} and 8.67 refers to the solar abundance value \citep{2004Asplund}.

We 
adopt
a sample-wide oxygen abundance using the
[NII]/H$\alpha$
values from \citet{2019Pereira-Santaella}. Here, we assume we are sampling the same population as \citet{2014Rigopoulou} and \citet{2014Magdis}, and thus these galaxies share similar chemical abundances. 
The global [NII]/H$\alpha$ value was converted to an oxygen abundance using the \citet{2004PettiniPagel} relation.

We obtain a conversion factor of $\alpha_\text{CO(2-1)} = 2.3 \pm 1.9~M_\odot(\text{K~km~s}^{-1}\text{~pc}^2)^{-1}$. Although our choice in adopted oxygen abundance could impact the final result, our $\alpha_\text{CO(2-1)}$ falls between the $\alpha$ values for local ULIRGs and $z \sim 1.5$ ULIRGs \citep[$\alpha=3.6 \pm 0.8$;][]{2010DaddiA}, indicating our assumptions are reasonable. The final molecular gas masses for this sample following equation~(\ref{eq:gas_mass}) are in Table~\ref{tbl:phys_props}.

An additional systematic uncertainty may arise if the metallicities of individual galaxies differ from the sample-wide abundance adopted here. This may be particularly relevant in interacting systems, where accretion of lower-metallicity material from a companion or radial gas flows could modify the central gas-phase abundance. Locally, galaxy pairs have been observed to have a lower metallicity compared to an isolated counterpart with the same stellar mass \citep{2006Kewley, 2008Ellison, 2012Scudder, 2020Bustamente}, with preliminary indications of a similar pattern at high-redshift \citep{2019Wilson, 2021Horstman}. Because the adopted $\alpha_\text{CO(2-1)}$ is metallicity dependent, such variations would propagate directly into the inferred $M_\text{gas}$. In the absence of homogeneous object-by-object metallicity measurements, we do not attempt to account for this effect individually and regard it as a systematic uncertainty in the absolute $M_\text{gas}$.

Figure~\ref{fig:luminosity_mass} plots the infrared luminosity against molecular gas mass. Our sample falls between warm ULIRGs and cool LIRGs at the same redshift \citep{2011Combes,2017Lee}. The molecular gas masses of these 
diverse U/LIRG populations vary by up to a factor of 10. The intermediate-$z$ U/LIRGs have $M_\text{gas}$ values spreading across $\log(M_\text{gas}/M_\odot)\sim 10.1-11.5$. 
For comparison, a local survey of star-forming galaxies is included, with a notably lower mass range \citep{2018Cheng}.

\subsection{Surface densities and star formation efficiencies}

We next turn to characterize the star formation nature of this sample, examining primarily the star formation efficiency (SFE) and gas depletion timescale $\tau$. To do so, we use the molecular gas masses derived previously and information gleaned from each galaxy's photometry.
Using the BC03 stellar population synthesis code \citep{2003BruzualCharlot}, we have generated simple models with
exponentially decaying and/or constant star formation histories to fit the SEDs of our sample ULIRGs and
estimate their stellar masses.
We find that these values together with the effective radii produce characteristic velocities broadly consistent with our observed CO spectra FWHMs, following the virial theorem. A more thorough analysis of the kinematics of these galaxies is underway to produce a robust determination of the dynamical mass and rotational and dispersion velocities.
We also note that the errors associated with these models reflect the statistical uncertainties from the fitting process itself and not the systematic uncertainties coming from assumptions required to initialize the models, such as the physical and stellar properties of each object.

The molecular mass surface density was taken as the ratio between the molecular gas mass and the measured star-forming region: $\Sigma_\text{mol} = M_\text{gas}/A_\text{CO}$. Similarly, the star formation surface density is defined as $\Sigma_\text{SFR} = \text{SFR}/A_\text{CO}$. 
We perform this calculation only for targets visible in the 3$\sigma$ moment 0 map, defining  $A_\text{CO}$ as the product of the number of pixels in the map and the pixel area.

The star formation efficiency (SFE) is calculated as $\mathrm{SFE} = \Sigma_\text{SFR}/\Sigma_\text{mol}$. As we assume the emitting areas are the same for both surface densities, the SFE essentially becomes $\mathrm{SFE} = \mathrm{SFR}/M_\text{gas}$. The SFE values of this sample occupy the space between MS-like LIRGs \citep{2017Lee} and SB-like warm ULIRGs \citep{2011Combes} at this same redshift (Figure~\ref{fig:luminosity_SFE}). For the \citet{2011Combes} sample, we converted the SFE from $L_\odot/M_\odot$ to Myr by dividing by $5.8 \times 10^3$, following their analysis. We derived the SFE for the \citet{2016Decarli}, \citet{2018Cheng}, and \citet{2022Lamperti} samples by taking the ratio of the SFR to the dust mass. With the latter two samples, we used the \citet{2011Murphy} relation to convert the provided $L_\text{IR}$ to SFR. 

\begin{figure}
	\includegraphics{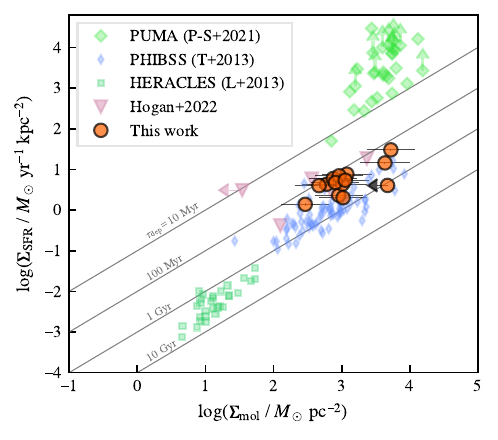}
    \caption{Star formation rate surface density over molecular gas mass surface density. Gray lines indicate different depletion timescales from 10~Myr to 10~Gyr. The average $\tau_{\rm dep}$ of this study is around hundreds of Myr, between $\tau_{\rm dep}\sim 1$~Gyr of local and $1<z<3$ MS galaxies in squares and diamonds, respectively \citep{2013Leroy,2013Tacconi}, and $\tau_{\rm dep}\lesssim 10$~Myr of highly interactive local ULIRGs as crosses \citep{2021Pereira-Santaella}. Also included is a smaller sample of intermediate-$z$ ULIRGs marked as triangles \citep{2022Hogan}.}
    \label{fig:surface_density}
\end{figure}

The ratio between the star formation surface density and the molecular mass surface density gives the characteristic depletion time $\tau=\Sigma_\text{SFR}/\Sigma_\text{mol}$, which roughly provides a timescale for the conversion of gas to stars. Figure~\ref{fig:surface_density} plots the two surface densities against each other, with solid lines denoting various time depletion scales. Our  sample of ULIRGs generally utilizes their gas supply on a shorter timescale of about 100~Myr. For the \citet{2021Pereira-Santaella} sample, we converted the $L_\text{IR}$ surface density to $\Sigma_\text{SFR}$ using the \citet{2011Murphy} relation.

The derived surface densities should be interpreted with caution because the CO emission is only marginally resolved in some targets and may contain a substantial unresolved central component. In such cases, the measured $A_\text{CO}$ does not necessarily represent the intrinsic $M_\text{gas}$ surface density-weighted area, thus introducing a systematic uncertainty into the absolute values of both $\Sigma_\text{mol}$ and $\Sigma_\text{SFR}$. We therefore use these quantities primarily for comparison with previous studies. Importantly, because we adopt the same $A_\text{CO}$ for the molecular gas and star-forming regions, the area eventually cancels when calculating the SFE, such that $\mathrm{SFE} = \mathrm{SFR}/M_\text{gas}$ and $\tau=M_\text{gas}/\mathrm{SFR}$. The depletion timescales discussed here are therefore independent of the adopted $A_\text{CO}$ under this assumption. 

We follow up on discussion of Figures~\ref{fig:luminosity_mass}--\ref{fig:surface_density} presented here in Section~\ref{sec:discussion}. All physical properties are reported in Table~\ref{tbl:phys_props}.

\subsection{The location on the MS-SB plane}

The classification of a star-forming galaxy relative to an MS galaxy has been classically defined by $\text{sSFR} = \text{SFR}/M_\star$ \citep{2004Brinchmann,2012Whitaker,2015Schreiber,2023Popesso}. From empirical data, numerous relations characterizing the slope and scatter of the MS correlation have been derived, from which a galaxy's distance from the MS can be defined \citep[e.g.,][]{2007Daddi,2007Elbaz,2014Speagle,2014Whitaker,2015Tasca,2019Popesso}.
However, $M_\star$ is highly variable depending on how modeling parameters (like stellar population models) are set for SED fits. 
Instead, we first consider the correlations based on 
the directly observable parameters $L_\text{IR}$ and $L'_\text{CO(1--0)}$ \citep{2014Sargent}.

The CO(2--1) luminosity was converted to a CO(1--0) luminosity $L'_{\text{CO(1--0)}}$ using the conversion factor $R_{21}=L'_{\text{CO(2--1)}}/L'_{\text{CO(1--0)}}=0.75$, which is derived from intermediate-$z$ ULIRGs \citep[][and references therein]{2014Magdis}

\begin{figure*}
	\includegraphics{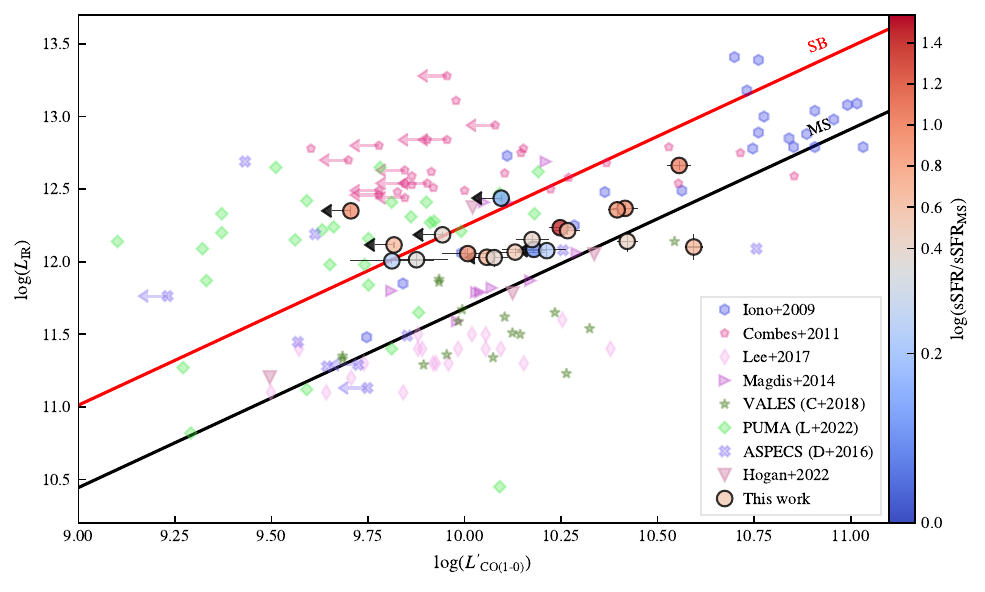}
    \caption{Infrared luminosity over CO line luminosity, including samples from the local Universe to high redshift. The \citet{2014Sargent} relations for MS and SB are plotted in thick lines and labeled. The intermediate-$z$ ULIRG sample is outlined in black and color-coded according to each galaxy's deviation above its corresponding \citet{2014Speagle} MS level for its given $M_\star$ and age, in dex. Both diagnostics affirm that the present sample lies in the region between MS and SB galaxies. Other populations represented here are discussed in Section~\ref{sec:add_samples} \citep{2011Combes,2014Magdis,2016Decarli,2017Lee,2018Cheng,2022Lamperti,2022Hogan}. A color version is available in the online journal.}
    \label{fig:LIR_LCO}
\end{figure*}

In Figure~\ref{fig:LIR_LCO}, we plot $L_\text{IR}$ over $L'_\text{CO(1--0)}$ with the \citet{2014Sargent} MS and SB designations in solid lines with labels. Our ULIRGs sample is color-coded according to each object's deviation above the MS defined by \citet{2014Speagle} in dex. The equation derived in that study provides the SFR of a standard MS galaxy based on the provided $M_\star$ and age. We compare the \citet{2014Speagle} MS SFR for each galaxy with our own derived SFR to determine the deviation from the MS. The two definitions for the MS parameter space generally agree for our sample --- both methods place our sample in the region between MS and SB.

Different conversion factors were applied to samples from the literature to produce $L'_\text{CO(1--0)}$. $L'_\text{CO(3--2)}$ values for U/LIRG samples were converted to $L'_\text{CO(1--0)}$ using $R_{31}=0.47$ \citep{2010Leech}. To convert the $L'_\text{CO(3--2)}$ values for the \citet{2022Lamperti} sample, we used $R_{21}=0.91$, derived from local LIRGs \citep{2012Papadopoulos}.
We applied $R_{21}=0.64$ to the \citet{2018Cheng} sample based on a study of nearby star-forming galaxies \citep{2021denBrok}. The \citet{2014Magdis} $L'_\text{CO(1--0)}$ were taken as cited in their publication.

\subsection{Evolution with redshift} \label{sec:analysis-redshift}
\begin{figure*}
	\includegraphics{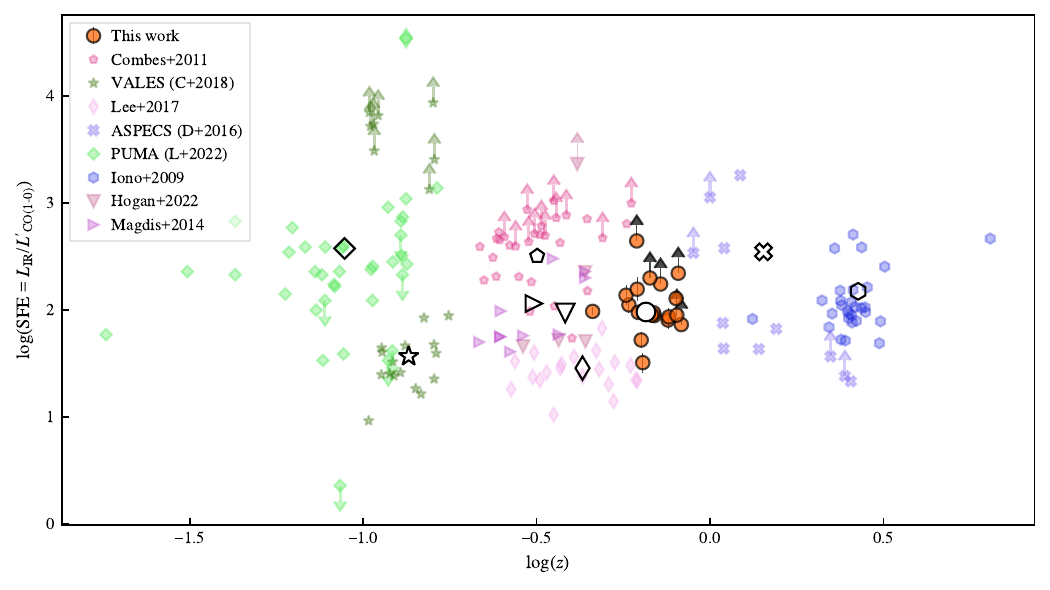}
    \caption{Star formation efficiency over redshift, where SFE is defined as the ratio of infrared to CO(1--0) line luminosity. Across redshift, the survey presented here occupies a region distinct from known MS and SB galaxies. Further details on other surveys from the literature are compiled in Section~\ref{sec:add_samples} \citep{2009Iono,2011Combes,2014Magdis,2016Decarli,2017Lee,2018Cheng,2022Lamperti,2022Hogan}. The mean value of each sample, excluding upper limit estimations, are featured in larger, black-outlined markers.}
    \label{fig:SFE_redshift}
\end{figure*}

In considering our sample's 
evolutionary history, 
we opt to compare direct observables to remove the bias from additional assumptions (e.g., varied methods for deriving $\alpha$). We expand on the analysis from the previous section and examine $\text{SFE} = L_\text{IR}/L'_\text{CO}$ over redshift. This proxy for SFE is based on $L_\text{IR}$ tracing the SFR and $L'_\text{CO}$ tracing the molecular gas mass. 
As before, we convert all measurements 
to CO(1--0) line luminosities.
The resulting plot is shown in Figure~\ref{fig:SFE_redshift}. The mean of each sample is plotted in larger, clear, black-outlined markers. The mean of the present sample is shown with a filled white circle for clarity. 
It is evident that our sample of intermediate redshift ULIRGs shows lower SFE compared to the sample of local ULIRGs \citep{2022Lamperti} or the warm ULIRGs of \citet{2011Combes}. Instead, our intermediate redshift ULIRGs have more similar SFE to high redshift MS galaxies.

\begin{landscape}
	\begin{table}
		\centering
		\footnotesize
		\caption{CO line properties for the sample. From left to right: truncated object name, integrated flux density, CO line luminosity, detected line FWHM, molecular gas mass, molecular gas mass surface density, SFR surface density, star formation efficiency, and depletion timescale. Upper limits are provided for non-detections.\label{tbl:line_props}}

		\begin{tabular}{@{}lccccccccc@{}}
		\hline
		Object        & $S_\text{CO} \Delta v$ & $L'_\text{CO(2-1)}$                                  & FWHM    & $M_\text{gas}$    & $\log  \Sigma_\text{mol} $ & $\log \Sigma_\text{SFR}$  & \multicolumn{1}{c}{SFE}        & $\tau_\text{dep}$ \\
              & $\left(\text{Jy km s}^{-1}\right)$       & $\left( 10^9~\text{K km s}^{-1}\text{ pc}^2 \right)$ & $\left(\text{km s}^{-1}\right)$ & $\left(10^{10}~M_\odot \right)$ &      $\left( M_\odot~\text{pc}^{-2} \right)$                                    &         $\left( M_\odot~\text{yr}^{-1}~\text{pc}^{-2} \right)$                              & \multicolumn{1}{c}{$\left(\text{Gyr}^{-1}\right)$} & $\left(\text{Myr}\right)$               \\ \hline
			  	HELP\_J021636 & $1.30 \pm 0.25$ & $5.6 \pm 1.1$    & 419  & $1.3 \pm 1.1$   & $2.78 \pm 0.36$ & $0.636 \pm 0.075$ & $7.2 \pm 6.1$ & $140 \pm 119$ \\
				HELP\_J021642 & $1.70 \pm 0.26$ & $7.6 \pm 1.2$    & 417  & $1.8 \pm 1.5$   & $2.67 \pm 0.36$ & $0.600 \pm 0.064$ & $8.6 \pm 7.2$ & $117 \pm 98$  \\
				HELP\_J021654 & $3.31 \pm 0.18$ & $18.6 \pm 1.0$   & 205  & $4.4 \pm 3.6$   & $2.91 \pm 0.35$ & $0.677 \pm 0.057$ & $5.9 \pm 4.8$ & $170 \pm 140$ \\
				HELP\_J021742 & $3.19 \pm 0.25$ & $19.5 \pm 1.5$   & 494  & $4.6 \pm 3.7$   & $3.05 \pm 0.35$ & $0.736 \pm 0.033$ & $4.8 \pm 3.9$ & $207 \pm 170$ \\
				HELP\_J021814 & $1.48 \pm 0.28$ & $11.3 \pm 2.2$   & 564  & $2.7 \pm 2.2$   & $3.64 \pm 0.36$ & $1.160 \pm 0.027$ & $3.3 \pm 2.8$ & $299 \pm 250$ \\
				HELP\_J021820 & $<0.93$         & $<6.57$          & -    & $<15.44$        & -               & -                 & $<0.81$       & $<1240$       \\
				HELP\_J021824 & $4.85 \pm 0.33$ & $13.26 \pm 0.90$ & 447  & $3.1 \pm 2.5$   & $2.88 \pm 0.35$ & $0.777 \pm 0.030$ & $8.0 \pm 6.5$ & $126 \pm 103$ \\
				HELP\_J021856 & $3.73 \pm 0.23$ & $19.8 \pm 1.2$   & 314  & $4.6 \pm 3.8$   & $2.96 \pm 0.35$ & $0.366 \pm 0.078$ & $2.6 \pm 2.1$ & $390 \pm 325$ \\
				HELP\_J021929 & $<0.84$         & $<4.92$          & -    & $<11.56$        & -               & -                 & $<1.2$        & $<852$        \\
				HELP\_J021941 & $1.32 \pm 0.23$ & $10.1 \pm 1.8$   & 288  & $2.4 \pm 2.0$   & -               & -                 & $4.8 \pm 4.0$ & $208 \pm 174$ \\
				HELP\_J022057 & -               & -                & -    & -               & -               & -                 & -             & -             \\
				HELP\_J022117 & $<1.06$         & $<8.95$          & -    & $<21.02$        & -               & -                 & $<0.44$       & $<2267$       \\
				HELP\_J022127 & $<0.98$         & $<3.81$          & -    & $<8.94$         & -               & -                 & $<2.4$        & $<411$        \\
				HELP\_J022130 & $<1.07$         & $<9.33$          & -    & $<21.91$        & $<3.67$         & $0.607 \pm 0.022$ & $<0.86$       & $<1166$       \\
				HELP\_J022131 & $2.20 \pm 0.22$ & $11.2 \pm 1.1$   & 224  & $2.6 \pm 2.2$   & $3.01 \pm 0.35$ & $0.636 \pm 0.048$ & $4.2 \pm 3.5$ & $238 \pm 196$ \\
				HELP\_J022151 & $2.19 \pm 0.16$ & $13.9 \pm 1.0$   & 225  & $3.3 \pm 2.7$   & $2.47 \pm 0.35$ & $0.142 \pm 0.044$ & $4.7 \pm 3.9$ & $211 \pm 173$ \\
				HELP\_J022209 & $0.97 \pm 0.24$ & $4.9 \pm 1.2$    & 322  & $1.14 \pm 0.97$ & $3.08 \pm 0.37$ & $0.883 \pm 0.028$ & $6.4 \pm 5.5$ & $156 \pm 133$ \\
				HELP\_J022222 & -               & -                & -    & -               & -               & -                 & -             & -             \\
				HELP\_J022247 & $1.36 \pm 0.17$ & $8.6 \pm 1.1$    & 172  & $2.0 \pm 1.7$   & $3.72 \pm 0.36$ & $1.482 \pm 0.028$ & $5.7 \pm 4.7$ & $174 \pm 143$ \\
				HELP\_J022333 & $<1.30$         & $<12.24$         & -    & $<28.77$        & -               & -                 & $<0.36$       & $<2741$       \\
				HELP\_J022427 & $3.17 \pm 0.22$ & $27.0 \pm 1.9$   & 330  & $6.3 \pm 5.2$   & $2.96 \pm 0.35$ & $0.844 \pm 0.059$ & $7.6 \pm 6.3$ & $131 \pm 108$ \\
				HELP\_J022722 & $5.42 \pm 0.30$ & $29.4 \pm 1.6$   & 385  & $6.9 \pm 5.6$   & $3.02 \pm 0.35$ & $0.30 \pm 0.18$   & $1.9 \pm 1.8$ & $522 \pm 477$ \\ \hline
			  
		\end{tabular}
	\end{table}
\end{landscape}

\section{Discussion} \label{sec:discussion}

\subsection{Molecular gas mass and SFR}

We 
compare the present ULIRG sample with other U/LIRG surveys at intermediate redshift. Our primary samples of comparison are the \citet{2011Combes} warm ($>42$~K) interacting ULIRGs,  
the cooler LIRGs from \citet{2017Lee} that are mostly isolated disks, 
and the preliminary study of intermediate-$z$ ULIRGs in the northern hemisphere \citep{2022Hogan}, that motivated the present
investigation  as described in Section~\ref{sec:add_samples}.

Comparing
the SFE of these surveys in Figure~\ref{fig:luminosity_SFE} 
confirms the trend between $L_\text{IR}$ and SFE as noted by \citet{2022Hogan}. 
Our sample lies in a similar position to the \citet{2022Hogan} sample, between the lower luminosity and lower SFE LIRGs and the higher luminosity and higher SFE ULIRGs. U/LIRG samples at other redshifts generally follow these expected trends. 
The local ULIRGs from \citet{2022Lamperti}  
tend to have a higher SFE than the present sample, clustering closer to the \citet{2011Combes} ULIRGs. The \citet{2016Decarli} high-$z$ U/LIRGs 
instead reside more in the lower SFE region, near the \citet{2017Lee} sample.

Combining these SFR and gas mass values shows that the present sample does occupy an area distinct from MS and SB galaxies. The surface density plot in Figure~\ref{fig:surface_density} includes lines denoting different $\tau_\text{dep}$ values. The upper limits derived in this figure 
are estimates from the non-detections in the sample. Our sample has a $\tau_\text{dep}$ value on the order of hundreds of Myr, distinguishing it from MS galaxies with higher $\tau_\text{dep}$ on the order of 1~Gyr. For context, we include $1<z<3$ MS galaxies \citep{2013Tacconi} and local MS galaxies \citep{2013Leroy}. In comparison, our sample 
has a $\tau_\text{dep}$ value which is several times higher than the $\tau_\text{dep}<10~\text{Myr}$ of local ULIRGs \citep{2021Pereira-Santaella}. 
We thus conclude that our intermediate-$z$ ULIRGs appear to have a large molecular gas reservoir available for star formation, but they are consuming the gas at a slower pace than SBs, assuming negligible effects from beam size differences between our sample and others.

The physical properties of these galaxies 
and their location in these
plots
reveal their nature as ``transitioning'' objects not quite aligning with either typical MS values or SB-like values. As such, there could be a more gradual change in SFR among star-forming galaxies instead of the bimodality often observed, as suggested by \citet{2019Shangguan} and \citet{2022Hogan}. We further delve into the MS-SB classification itself to characterize 
their nature
better.

\subsection{MS-SB classification}

Assessing our sample of intermediate-$z$ ULIRGs within the sSFR-based and the observationally-based MS-SB relations,
we confirm that 
the sample does straddle the MS-SB transition (Figure~\ref{fig:LIR_LCO}). Following the \citet{2014Speagle} MS relation, we find our targets 
typically
lie within a dex above the MS, 
indeed in that region between MS and SB. 
Our sample also falls between the MS- and SB-defined regions in $L_{\rm {IR}}$ vs. $L'_{\rm{CO(1-0)}}$ space, following the \citet{2014Sargent} correspondence.
Although these two relations do not strictly agree on a galaxy-by-galaxy basis, they both confirm the overall trend of transition we had predicted. 

Indeed, we find the \citet{2014Sargent} relation robust at distinguishing MS and SB populations when we extend our analysis to include other samples. Local star-forming galaxies from the VALES survey hug the lower MS line \citep{2018Cheng}. Local ULIRGs from the PUMA survey extend well above the SB line \citep{2022Lamperti}. The comparative intermediate-$z$ U/LIRG samples also find alignment with this observable-based correlation---the warm ULIRGs from \citet{2011Combes} rise above the SB delineation, and the cold LIRGs from \citet{2017Lee} tightly follow the MS relation.

\subsection{SFE and distance above the MS} \label{sec:disc-sfe}

We unite the previous two sections by examining the 
evolution of
$\text{SFE}=L_\text{IR}/L'_\text{CO}$ over redshift in Figure~\ref{fig:SFE_redshift}. 
MS-like samples have lower SFE, and SB-like samples occupy the higher parameter space, in good agreement with their classification based on $\mathrm{SFE} = \Sigma_\text{SFR}/\Sigma_\text{mol}$ and sSFR.
As seen previously, our sample falls in a distinct region between MS and SB samples at both low and intermediate redshift. The high redshift MS populations tend to have a slightly higher SFE than lower redshift MS samples, at a similar level to our intermediate-$z$ sample.

We caution that the distribution in Figure~\ref{fig:SFE_redshift} reflects the selection criteria outlined in Section~\ref{sec:sample_selection}. The sample is selected within the ULIRG regime and therefore spans a comparatively restricted range in $L_\text{IR}$, while the CO observations impose a sensitivity-dependent lower limit on detectable $L'_\text{CO}$.  At fixed $L_\text{IR}$, this can preferentially exclude CO-faint systems with high $L_\text{IR}/L'_\text{CO}$ values. Nevertheless, within these selection limits, the CO-detected sources clearly occupy an intermediate region relative to the comparison samples considered here.

Interestingly, the \citet{2018Cheng} sample occupies a different space among MS and SB samples based on SFE definition. With the observable definition of $\text{SFE}=L_\text{IR}/L'_\text{CO}$, the targets have values lying between those of the present sample and the \citet{2017Lee} LIRGs (Figure~\ref{fig:SFE_redshift}). In Figure~\ref{fig:luminosity_SFE}, however, the \citet{2018Cheng} objects have noticeably higher $\mathrm{SFE} = \Sigma_\text{SFR}/\Sigma_\text{mol}$, more in the range of SBs. 
The shift in SFE reveals the significant role that choice in conversion factors can play in affecting resulting physical quantities.

Overall, it appears that SFE is a good indicator of a galaxy's deviation from the MS, and in effect, its SFR enhancement. We find this result in agreement with previous studies on star formation indicators that have linked SFE to SFR enhancement \citep{2014Sargent,2016Saintonge,2019Shangguan,2020Piotrowska}. 
More explicitly, SFE has been suggested as a driver of SFR variation above and below the MS definition \citep{2016Saintonge,2020Ellison}. 

The gas fraction $f_\text{gas}=M_\text{mol}/(M_\text{mol} + M_\star)$ has also been discussed in connection with the sSFR. 
$f_\text{gas}$ has been strongly suggested to follow trends in SFR evolution, based on both local studies \citep{2013Combes} and high-redshift observations and simulations \citep{2017Fensch,2018Elbaz}.
Although we do not consider $f_\text{gas}$ here, as we focus on analyzing this sample using direct observables, we cannot rule out a potential interplay between SFE, $f_\text{gas}$, and SFR.

Having confirmed through multiple avenues the transitional nature of these objects in the MS-SB regime, we next attempt to 
further constrain their properties.

\subsection{Minor mergers}

It is well known that 
major mergers
are behind the prodigious 
infrared 
luminosities of 
ULIRGs \citep[e.g.,][]{1996MihosHernquist,2006Hopkins,2016Sparre}. 
The two merging galaxies in this process are commonly defined as having a mass ratio less than 3:1.
Broadly, 
the merger fraction has been shown to increase with distance above the MS \citep{2019Pearson,2019Cibinel}. Morphological studies of galaxies have connected these mergers to moderately enhanced star formation and even SBs \citep{2019Cibinel,2025Vasquez-Bustos}. \citet{2012Kartaltepe} found that roughly 50\% of $z \sim 2$ ULIRGs in their sample were interacting, a value that appeared to slightly increase with lower redshift upon examination of a $z \sim 1$ sample. Conversely, local U/LIRGs have almost exclusively been found to be major mergers 
\citep{1988Sanders}. 

The Suprime-Cam images show that roughly half the objects in our sample display disturbed morphologies---we see evidence of tidal tails, potential satellite galaxies, and interestingly in the case of HELP\_J021824.031-045304.606, what appears to be double nuclei. Although by eye the two objects seem to have similar mass, indicating a major merger, we lack additional information to confirm their physical proximity. 
Although we cannot rule out the role of major mergers or other evolutionary tracks entirely based on the implications from the Suprime-Cam images and the discovery of two potential sources within the HELP\_J021742.522-053700.041 system (Figure~\ref{fig:twosource}), higher resolution imaging is required to fully confirm this. Overall, however, our intermediate-$z$ ULIRGs seem to align more closely with the high redshift trend in interaction rate.
Consideration of the line kinematics will further expose the nature of interactions in this sample, which will be explored in future work with a more in-depth analysis of the sample morphologies. Currently, preliminary results do support the $\sim 50\%$ disturbance rate in the sample (Kim et al. 2026, in prep).

In addition to the morphologies, the line spectra in Figure~\ref{fig:spectra2} do not show obvious signatures of major mergers or disturbances. The profiles are largely centrally peaked, consistent with the concentrated molecular gas emission implied by the moment 0 maps (Figure~\ref{fig:3sig_mom0}). We caution, however, that the profile alone does not establish the spatial concentration of the molecular gas. A few galaxies (e.g., HELP\_J021941.586-044426.974 and HELP\_J022151.541-025349.283) do have hints of a double-peaked emission line that could indicate disk-like gas rotation. Again, ongoing kinematics work will help shed light on the motion of gas within these objects.

Minor mergers (typically with mass ratios greater than 3:1) have been argued to play a large role in enhancing star formation.
Model simulations of minor mergers with gas-abundant parent disks find 35-45\% of gas in the disk galaxy ended up collapsed in its center over a few hundred Myrs \citep{1989Hernquist,1995HernquistMihos}.
Another simulation finds that up to $z \sim 3.7$, the SFR enhancement induced by minor mergers is comparable to that of major mergers, bumping up the SFR by an average factor of 1.69 at cosmic noon and increasing with lower redshift \citep{2017Martin}.
Observations have connected minor mergers to increases in star formation,
with an average increase of $\sim0.3$~dex in SFR above the MS relation noted among minor (and major) mergers in disk-like star-forming galaxies \citep{2015Willett}. 
Locally, minor mergers have been found to drive $\sim 65\%$ of LIRGs, implying they are capable of triggering star formation rates at the level of tens to hundreds of $M_\odot~\text{yr}^{-1}$ \citep{2015Carpineti}.
Such evidence validates previous claims of minor mergers inducing starbursts based on kinematics, morphological studies, and integral field spectroscopy \citep{2006Mazzuca,2012Alonso-Herrero}.
Broadly, minor mergers trigger roughly half of all local star formation activity, and their role is assumed to be greater at higher redshift with increasing merger rates and gas fractions \citep{2014Kaviraj}. In fact, major mergers have been seen to contribute to only one third of total mass growth at high redshift \citep[][and references therein]{2017Fensch}. At this intermediate redshift, minor mergers were found to occur three times more frequently than major mergers, with little change over redshift \citep{2011Lotz}. We thus turn to minor mergers as the 
potential driver of the
transitional tendencies in our sample.

As discussed in Section~\ref{sec:analysis-mgas}, the effects of an interaction can be impressed upon the overall metallicity of a system.
Metallicity dilution in merger pairs have been replicated in simulations \citep{2012Torrey, 2018Bustamente}. These studies have found minor mergers to be capable of having a relatively depressed metallicity, although to a lesser extent than that of major mergers.
Given the metallicity-dependence of $\alpha_\text{CO}$ \citep[][and references therein]{2015Genzel}, the dilution effects from a minor merger could be carried through to affect molecular gas mass values and SFEs. 
We caution, however, that our results may not fully reflect the influence of a minor merger due to the lack of individual metallicity abundance measurements available for this sample.

Reconciling our discussion of minor mergers with our observed trend of SFE tracing star formation, we note some nuances in directly connecting these phenomena. First, the SFE may not immediately reflect merger effects. In simulations, a lag has been noted between a minor merger occurring (based on morphological evidence) and the resulting bump in SFE on an order of tens of Myr later.
This increase in SFE was shown to be a reflection of the increase in sSFR, elevating the galaxy to a higher position within the MS scatter before potentially dropping down again \citep{2022HoganThesis}. These findings are consistent with local studies: \citet{2019Shangguan} links high SFE LIRGs with no major merger signatures to minor mergers, and \citet{2025Vasquez-Bustos} identifies potential minor mergers that boost the SFE in the host galaxy within the MS regime.
This correlation between sSFR and SFE may be constrained to smaller timescales, however, as \citet{2022Renaud} found the correlation disappeared as a galaxy shifted evolutionary phases (e.g., from a starburst phase to a secular phase). Considering the tension between the classical definition of the MS with other indicators of star formation phase, then, more care must be taken when interpreting a galaxy's position in MS parameter space. Based on our current analysis,  
we suggest that
the transitional nature of our sample in the MS parameter space could be due to minor mergers occurring in the galaxy and temporarily increasing the sSFR, which is subsequently reflected in the SFE. Further analysis of this potential scenario will be presented in our forthcoming work on describing the kinematics of our intermediate-z ULIRGs.

\section{Conclusions} \label{sec:conclusions}
We present CO~(2--1) ALMA observations  of 22 intermediate redshift ($0.46<z<0.81$) ULIRGs. 
This is the most comprehensive sample to date of ULIRGs at this redshift, representing a complete range of dust temperatures ($25<T_{\rm dust}<50$~K). We summarize the main results here:
\begin{itemize}
	\item CO~(2--1) emission was detected in 14 out of 22 targets. Of those fourteen, roughly half appeared to show morphological disturbances based on examination of $g$- and $i$-band filter Suprime-Cam images. A detailed kinematics analysis using the CO line observations will be reported in a forthcoming study.
	\item Our analysis of the physical parameters of this sample confirm that our intermediate-$z$ ULIRG sample possesses characteristics distinct from MS- and SB-like galaxies. In particular, the SFEs of our sample fall in a unique range, indicating depletion timescales, $\tau_\text{dep}$, on the order of hundreds of Myr. This value falls between $\tau_\text{dep} \sim 1~\text{Gyr}$ noted in normal star-forming galaxies and $\tau_\text{dep}<10~\text{Myr}$ seen in SBs. 
	\item We find the SFE to be a strong tracer of SFR enhancement and indicator of position in MS parameter space. Examining the strict definition of SFE in conjunction with SFE estimates based on direct observables confirms that the relative SFE can be utilized to determine the intensity of star formation within a galaxy.
	\item In the present sample of intermediate redshift ULIRGs, we propose that minor mergers could be responsible for placing these galaxies in the transition region between MS and SB, although we cannot exclude the incidence of major mergers. Minor mergers could be efficient in bumping up the star formation (and subsequently, the SFE) and elevating galaxies temporarily from their initial position.
\end{itemize}

This representative sample of intermediate-$z$ ULIRGs falls between the bimodal distribution of MS and SB galaxies following multiple prescriptions of the Kennicutt-Schmidt relation, suggesting that the SFE is a strong indicator of the current mode of star formation in a galaxy.
Interactions, particularly minor mergers, appear to provide a viable route for producing this temporary enhancement although we cannot exclude other evolutionary pathways. Future analysis of the resolved molecular gas kinematics will help distinguish between the various scenarios and establish the kinematic properties of this sample.

\section*{Acknowledgments}

We thank the referee for their insightful and constructive comments in furthering the paper.

JHK is supported by the Clarendon Fund Scholarships, the Science \& Technologies Facilities Council Doctoral Training Partnership studentship [ST/Y509474/1], and the Mansfield-Jocelyn Bell Burnell UK Research and Innovation Partnership scholarship. DR acknowledges the support from a Leverhulme Trust Research Project Grant.
MPS acknowledges support from grants RYC2021-033094-I, CNS2023-145506, and PID2023-146667NB-I00 funded by MCIN\/AEI\/10.13039\/501100011033 and the European Union NextGenerationEU\/PRTR.
IGB is supported by the Programa Atracci\'on de Talento Investigador ``C\'esar Nombela'' via grant 2023-T1/TEC-29030 funded by the Community of Madrid, and acknowledges support from the research project PID2024-159902NA-I00 funded by the Spanish Ministry of Science and Innovation/State Agency of Research (MCIN/AEI/10.13039/501100011033) and FSE+.

This paper makes use of the following ALMA data: ADS\/JAO.ALMA\#2023.1.00424.S. ALMA is a partnership of ESO (representing its member states), NSF (USA) and NINS (Japan), together with NRC (Canada), MOST and ASIAA (Taiwan), and KASI (Republic of Korea), in cooperation with the Republic of Chile. The Joint ALMA Observatory is operated by ESO, AUI/NRAO and NAOJ.

\section*{Data Availability}
The ALMA data presented in this paper are available in the ALMA archive.

This paper is based in part on data from the Hyper Suprime-Cam Legacy Archive (HSCLA), which is operated by the Subaru Telescope. The original data in HSCLA were collected at the Subaru Telescope and retrieved from the HSC data archive system, which is operated by the Subaru Telescope and Astronomy Data Center at National Astronomical Observatory of Japan. The Subaru Telescope is honored and grateful for the opportunity of observing the Universe from Maunakea, which has the cultural, historical and natural significance in Hawaii.

This paper makes use of software developed for the Vera C. Rubin Observatory. We thank the observatory for making their code available as free software at http://dm.lsst.org.



\bibliographystyle{mnras}
\bibliography{main} 




\appendix

\section{Supplemental figures}

\begin{figure}
	\includegraphics{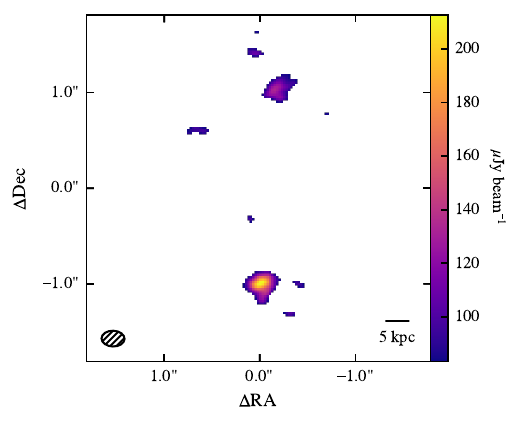}
    \caption{A zoomed out moment 0 map of HELP\_J021742.522-053700.041, showing the two sources present. We choose to focus our analysis on the brighter southern object.}
    \label{fig:twosource}
\end{figure}

\begin{figure*}
	\includegraphics{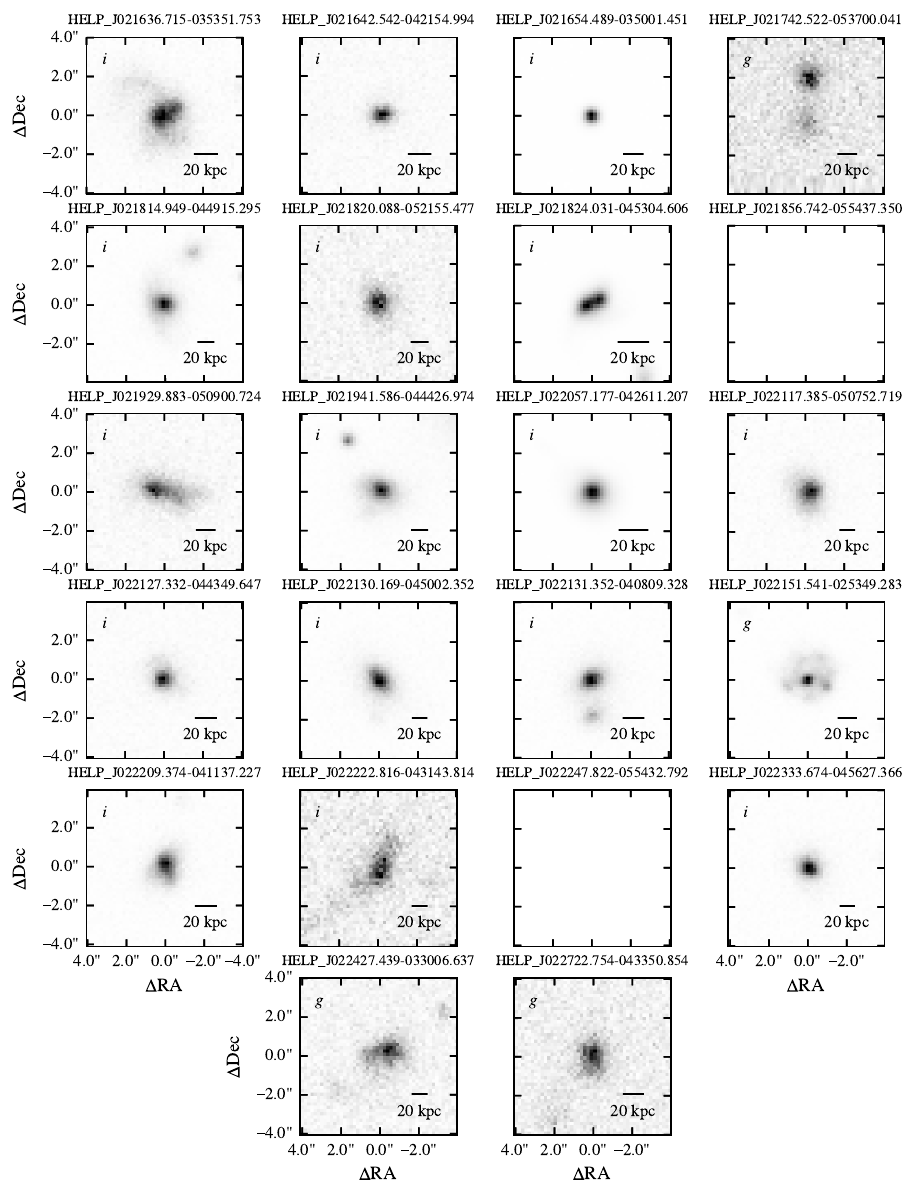}
    \caption{Images of the sample from the Suprime-Cam archive. Images were taken at either \textit{i} or \textit{g} bands based on availability, and are indicated as such in the top left corners. The grayscale is normalized to the brightest pixel in each cutout. Clear images for HELP\_J021856.742-055437.350 and HELP\_J022247.822-055432.792 were not available. The scale bar at the bottom right of each plot defines 20~kpc.}
    \label{fig:suprimecam}
\end{figure*}

\begin{figure*}
	\includegraphics{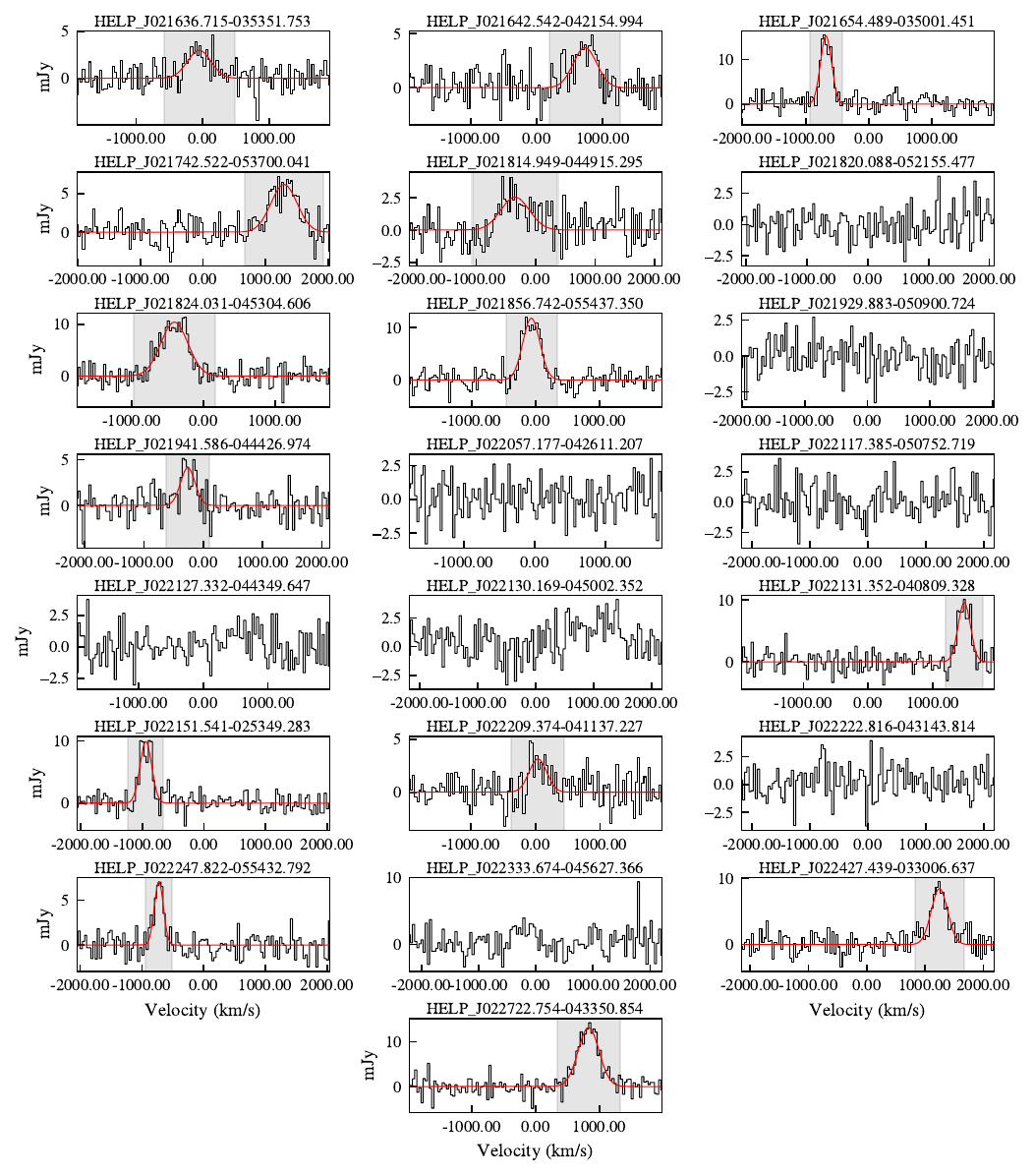}
    \caption{Spectra of the sample, taken with a 0.5\arcsec radius circular mask about the center of each object. Section~\ref{sec:analysis-comaps} details the spectra extraction method. Detected lines are fitted with a Gaussian. The gray bands cover 3$\sigma$ on either side of the observed central frequency, as defined by the fit. The central velocity on each x-axis is based on the initial spectroscopic redshift from HELP used to define the observed spectral window. Out of 22 total objects, 14 had a clear line detection.}
    \label{fig:spectra2}
\end{figure*}

\begin{table*}
	\centering
	
	\caption{Samples referenced in this work, including the redshift range of their objects and $L_\text{IR}$ range. More detailed descriptions of the samples can be found in Section~\ref{sec:add_samples}.\label{tbl:surveys}}
	\begin{tabular}{lcc}
	\hline
	Reference                                                         & $z$    & $\log (L_\text{IR}/L_\odot)$ \\ \hline
	HERACLES \citep{2013Leroy}                       & local        & -                                  \\
	PUMA \citep{2021Pereira-Santaella, 2022Lamperti} & $<0.165$     & 12.0--12.7                         \\
	VALES \citep{2018Cheng}                          & $<0.2$       & 10.72--12.14                       \\
	\citet{2011Combes}                               & 0.2--0.6     & 12.44--13.28                       \\
	\citet{2014Magdis}                               & 0.21--0.88   & 11.59--12.69                       \\
	\citet{2017Lee}                                  & 0.25--0.65   & 11.1--11.6                         \\
	\citet{2022Hogan}                                & 0.289--0.437 & 11.79--12.41                       \\
	This work                                                         & 0.46--0.81   & 12.01--12.66                       \\
	ASPECS \citep{2016Decarli}                       & 0.8--2.6     & 11.1--12.7                         \\
	PHIBSS \citep{2013Tacconi}                       & 1--3         & -                                  \\
	\citet{2009Iono}                                 & 2--3         & 10.99--13.93                       \\ \hline
	\end{tabular}
\end{table*}


\bsp	
\label{lastpage}
\end{document}